\documentclass[sigplan,10pt]{acmart}

\AtBeginDocument{%
  \providecommand\BibTeX{{%
    \normalfont B\kern-0.5em{\scshape i\kern-0.25em b}\kern-0.8em\TeX}}}

\graphicspath{{./images/}} 

\usepackage{tikz}
\usetikzlibrary{positioning, arrows, backgrounds, patterns, calc, shapes, fit}
\usepackage{listings}
\usepackage{xcolor}
\usepackage{tcolorbox}
\usepackage{colortbl}
\usepackage{graphicx}
\tcbuselibrary{listings,skins}
\usepackage{algorithm}
\usepackage{algpseudocode}   
\usepackage{amsmath}
\usepackage{multirow}
\usepackage{booktabs}
\usepackage{scalerel}
\usepackage{booktabs}
\usepackage[subtle]{savetrees}
\usepackage{hyphenat}
\usepackage{enumitem}
\setlist[itemize]{align=left, leftmargin=*}
\setlist[enumerate]{align=left, leftmargin=*}

\definecolor{initcode}{RGB}{213,232,212}    
\definecolor{cleancode}{RGB}{255,242,204}   
\definecolor{guideaccess}{RGB}{225,213,231} 
\definecolor{initoutline}{RGB}{130,179,102} 
\definecolor{cleanoutline}{RGB}{214,182,86} 
\definecolor{guideoutline}{RGB}{150,115,166} 

\newcommand{\ARC}{ATC}
\newcommand{\ARCfull}{Active Thread Count}

\lstdefinestyle{threadcode}{
  basicstyle=\scriptsize\ttfamily,
  columns=flexible,
  keepspaces=true,
  basewidth=0.5em,
  resetmargins=true,
  xleftmargin=0pt,
  xrightmargin=0pt,
  aboveskip=0pt,
  belowskip=0pt,
  lineskip=-2pt,
  boxpos=t,
  framesep=0pt
}

\newcommand{\myparagraph}[1]{%
  \noindent
  \textbf{
    #1%
    .
  }
}

\begin{document}

\title{Tidying Up the Address Space}


\newcommand{\PlainAuthors}{Vinay Banakar, Suli Yang, Kan Wu, Andrea C. Arpaci-Dusseau, Remzi H. Arpaci-Dusseau, Kimberly Keeton}

\renewcommand{\thefootnote}{\fnsymbol{footnote}}

\author{
  Vinay Banakar\textsuperscript{1, 3}, 
  Suli Yang\textsuperscript{3}, 
  Kan Wu\textsuperscript{2}\footnotemark[1],
  Andrea C. Arpaci-Dusseau\textsuperscript{1}, \\ 
  Remzi H. Arpaci-Dusseau\textsuperscript{1}, 
  Kimberly Keeton\textsuperscript{3} \vspace{1.5ex} \\ 
  \textsuperscript{1}University of Wisconsin-Madison \quad 
  \textsuperscript{2}xAI \quad 
  \textsuperscript{3}Google \\
}


\copyrightyear{2025}
\acmYear{2025}
\setcopyright{cc}
\setcctype{by}
\acmConference[DIMES '25]{3rd Workshop on Disruptive Memory Systems}{October 13--16, 2025}{Seoul, Republic of Korea}
\acmBooktitle{3rd Workshop on Disruptive Memory Systems (DIMES '25), October 13--16, 2025, Seoul, Republic of Korea}
\acmDOI{10.1145/3764862.3768179}
\acmISBN{979-8-4007-2226-4/25/10}



\renewcommand{\shortauthors}{ V. Banakar, Y. Suli, K. Wu, A. Arpaci-Dusseau, R. Arpaci-Dusseau, K. Keeton}

\begin{abstract}
Memory tiering in datacenters does not achieve its full potential due to hotness fragmentation—the intermingling of hot and cold objects within memory pages. 
This fragmentation prevents page-based reclamation systems from distinguishing truly hot pages from pages containing mostly cold objects, fundamentally limiting memory efficiency despite highly skewed accesses. 
We introduce address-space engineering: dynamically reorganizing application virtual address spaces to create uniformly hot and cold regions that any page-level tiering backend can manage effectively. 
HADES demonstrates this frontend/backend approach through a compiler-runtime system that tracks and migrates objects based on access patterns, requiring minimal developer intervention. 
Evaluations across ten data structures achieve up to 70\% memory reduction with 3\% performance overhead, showing that address space engineering enables existing reclamation systems to reclaim memory aggressively without performance degradation.
\end{abstract}






\maketitle

\footnotetext[1]{Work done at Google}



\section{Introduction}

Memory is the most constrained and costly resource in modern datacenters, with utilization reaching 60-90\% at hyperscalers like Google~\cite{google-borg} and Meta~\cite{tpp} while driving 50\% of server costs~\cite{pond}. To control these mounting expenses, operators adopt \emph{memory tiering} to move cold data from expensive DRAM to slower, cheaper storage~\cite{tpp, pond, memtis, tmts}. 
However, the effectiveness of tiering is fundamentally limited by a semantic gap between applications and the operating system. 
Applications operate on fine-grained data objects, while the OS manages memory in coarse-grained pages. 
This mismatch causes frequently accessed (hot) and infrequently accessed (cold) objects to become intermingled on the same pages—a problem we term \textit{hotness fragmentation}. 
This fragmentation cripples page-level reclamation, as even a single hot object on a page prevents the entire page from being safely moved to a slower tier without risking performance-degrading page faults.

Existing systems do not address the root cause of this problem. Page-level reclamation systems like the kernel's kswapd, Google's zswap~\cite{zswap}, or Meta's TMO~\cite{TMO} are powerful but ignorant; they can only identify pages as hot or cold, not the objects within them, and thus cannot distinguish a truly hot page from a mostly-cold one. Allocation-time hinting approaches~\cite{llvm-pgho} are too static, making a one-time placement decision that cannot adapt as an object's access patterns change over its lifetime. While more radical object-level tiering systems~\cite{aifm, mira} offer fine-grained control, they impose high adoption costs by requiring significant application modifications and direct hardware access. 
Therefore, what is needed is a system that provides object-level awareness to guide page-level decisions, without abandoning the abstractions that make existing systems practical.

\begin{figure}[!t]
  \centering
  \includegraphics[width=0.484\textwidth]{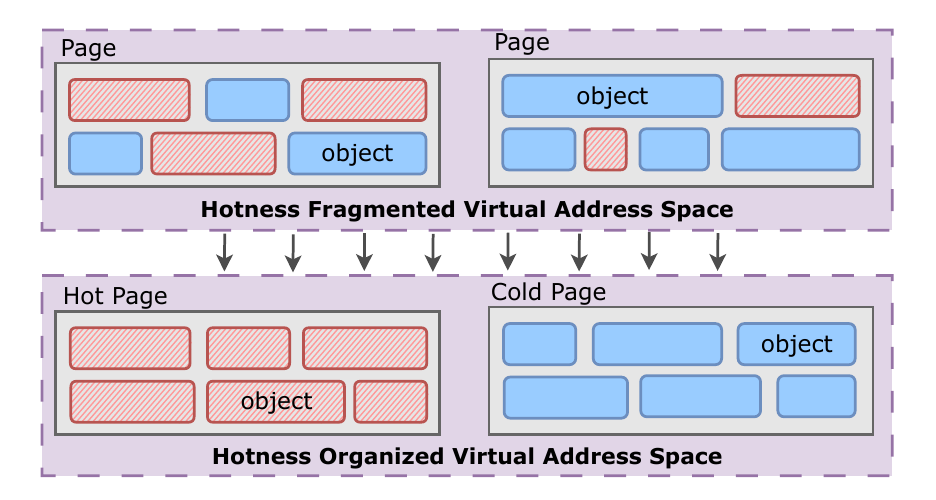}
  \caption{\textbf{Address-Space Engineering.} 
  \small{Hotness fragmentation (top) intermixes hot (red) and cold (blue) objects, making pages difficult to reclaim. Address-space engineering (bottom) reorganizes objects into uniformly hot and cold regions, enabling efficient page-level management.}}
  \label{fig:address-space-engineering}
\vspace*{-6mm}
\end{figure}

We propose \textbf{address-space engineering}: instead of making the OS object-aware, we engineer the application's virtual address space to be OS-tiering-friendly, adapting to workload access patterns.
While garbage-collected languages have long employed object reorganization for locality~\cite{oor-oopsla-2004, halo-pldi-2006, cache-conscious-2000, white-1980}, unmanaged languages like C++ present unique challenges due to the assumption that allocation-time decisions are final~\cite{mythGC}. The key insight is to challenge this assumption for pointer-based data structures while preserving the language's semantic guarantees.
As shown in Figure~\ref{fig:address-space-engineering}, this approach "tidies up" the fragmented address space by dynamically identifying and migrating objects based on their access patterns, transforming a fragmented layout, where hot and cold data are mixed, into a cleanly organized one, where objects are grouped by their access temperature. This continuous, dynamic engineering of the virtual address space is the key to bridging the semantic gap between application objects and OS pages.

Dynamic reorganization enables a powerful architectural principle: decoupling object-level placement from page-level reclamation. We propose a model with a frontend system responsible for making object placement decisions and a backend responsible for managing the underlying pages. The frontend provides the backend with an address space containing uniformly hot or cold regions, making the backend's job easier: it can confidently reclaim entire cold regions or promote hot regions to huge pages. This architecture achieves the intelligence of an object-level system while retaining the compatibility and simplicity of existing page-level systems.

In this paper, we present 
\textbf{\textit{Hierarchically} \textit{Aware} \textit{Data} \textit{structurES} \textit{(HADES)}},
a compiler-runtime system that realizes this frontend/backend vision for pointer-based data structures in non-managed languages. HADES transparently tracks object-level access intensity and uses a safe, lock-free protocol to migrate objects between hot and cold heaps. We demonstrate that by adding a HADES frontend to standard reclamation approaches like kswapd and proactive reclamation, we can reduce memory usage by up to 70\% on YCSB workloads with minimal performance degradation (3-5\%). This work makes the following contributions: (1) a decoupled frontend/backend model for memory tiering based on dynamic object reorganization; (2) the design and implementation of HADES, an object-level frontend; and (3) an evaluation demonstrating that this model allows existing, unmodified backends to reclaim memory far more effectively.
\vspace{-1mm}




\section{Semantic Gap in Memory Management}

Real-world workloads exhibit highly skewed access patterns, with large portions of datasets remaining untouched~\cite{benchmark-rocksdb-fb,first-gen-cxl} or accessed only once~\cite{s3-fifo}.
Modern allocators \cite{tcmalloc, jemalloc} optimize for allocation speed and spatial fragmentation reduction, making placement decisions at allocation time without considering future access patterns. 
Meanwhile, Linux's memory reclamation system operates at page granularity, using access bits in page table entries to track activity. 
This fundamental mismatch scatters frequently accessed objects across memory pages, intermixing them with rarely accessed data and creating a semantic gap between application-level object access and OS-level page management.

\begin{figure}[!t]
    \centering
    \includegraphics[width=0.484\textwidth]{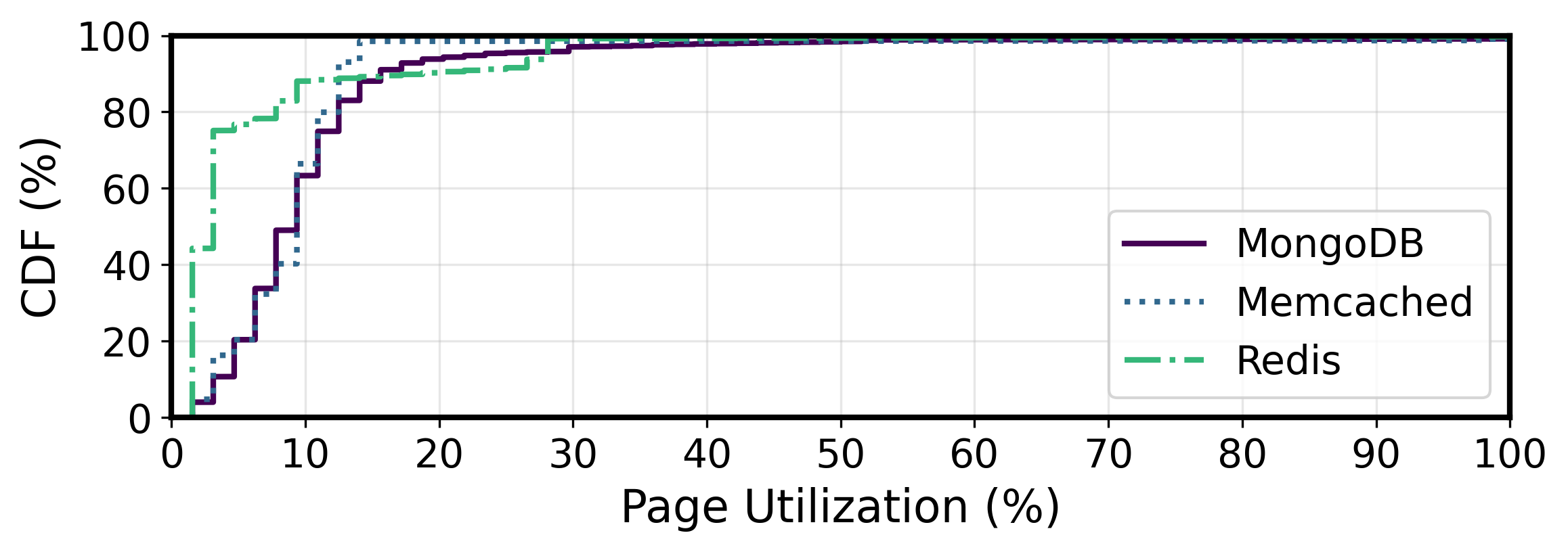}
    \caption{\textbf{Page Utilization.} 
    \small{Redis, Memcached, and MongoDB Page Utilization for 360s epochs running a YCSB-C (read-only) workload with a Zipfian distribution. A small fraction of bytes per page are accessed, and more pages are touched than necessary.}}
    \label{fig:redis-pin}
\vspace*{-6mm}
\end{figure}

From a memory tiering perspective, the core inefficiency arises from the fragmentation of large regions of cold data by a few hot objects, which makes entire pages unreclaimable. This intermingling of hot and cold objects within the same pages is a problem we term \textit{hotness fragmentation}. To quantify the extent of hotness fragmentation, we introduce the metric \emph{Page Utilization}. Page Utilization is defined as the ratio of unique accessed bytes to the total size of accessed pages over a specific time period T:
$$PageUtilization(T) = \frac{TotalUniqueBytes(T)}{UniquePages(T) \times PageSize}$$
A low Page Utilization value serves as a strong indicator of this problem: it signifies that only a small fraction of the bytes on an accessed page are actually being used. This forces the entire page to remain resident in the faster memory tier, trapping the large volume of co-located, un-accessed (and likely cold) data on the same page.

Our investigations reveal consistently low Page Utilization across diverse systems and workloads. 
Using PinTool~\cite{pintool} to track memory access patterns in popular key-value stores running YCSB workloads~\cite{ycsb}, we found that 75\% of accessed pages in Redis utilize just 3\% or less of their capacity, while 90\% of pages in MongoDB and Memcached use less than 15\% (Figure~\ref{fig:redis-pin}). 
This scattered placement creates an illusion of high memory activity when only a small fraction of bytes receives regular access, directly constraining the effectiveness of any page-based reclamation system.

The issue of non-uniform access within larger memory granularities is well-established. Cache prefetching studies~\cite{sfp, sms, bingo}, using techniques like spatial footprint bitmasks~\cite{sfp, bingo} and block access counts per region~\cite{sms}, have consistently revealed that only a fraction of data within multi-kilobyte granularities is often hot. These studies confirm that fine-grained access locality exists, but their block-level metrics are primarily designed to identify specific cache blocks for prefetching.

To assess the impact of this fine-grained skew on OS-level memory tiering, we use \emph{Page Utilization}. This metric quantifies the byte-level efficiency within the set of all unique OS pages touched, not just within a prefetch-specific region. By calculating the ratio of unique bytes accessed to the total size of these pages, Page Utilization directly measures the memory wasted due to "hotness fragmentation"—where sparsely located hot objects force entire pages to remain in the fast tier. Unlike metrics aimed at prefetch accuracy, low Page Utilization serves as a direct proxy for the reclaimability challenges faced by page-based memory management, highlighting the potential benefits of improving the access uniformity of pages.

\begin{figure}[!t]
    \centering
    \includegraphics[width=0.484\textwidth]{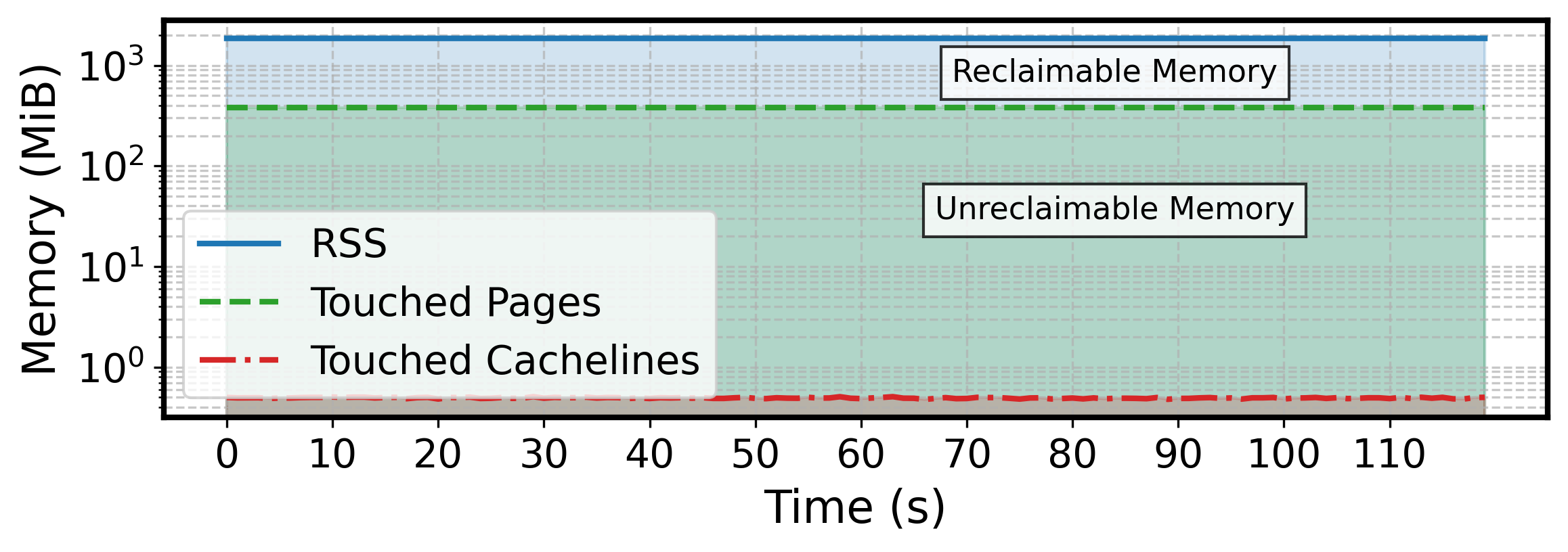}
    \caption{\textbf{Unreclaimable Memory.} 
    \small{YCSB-C with Zipfian distribution running on Redis shows memory used (RSS), pages needed (Touched Pages), and cachelines needed (Touched Cachelines). Only 0.5 MiB of actual data is required whereas 1.2 GiB remains resident. The gap represents theoretically reclaimable memory that current systems cannot efficiently recover.}}
    \label{fig:redis-reclaim}
\vspace*{-6mm}
\end{figure}

\paragraph{Penalties of Poor Page Utilization}
Memory reclamation in modern operating systems requires completely cold pages, as even a single active object prevents the entire page from being swapped out. 
Figure~\ref{fig:redis-reclaim} illustrates this problem with Redis running a YCSB-C workload, where despite requiring 1.2 GiB of resident memory, Redis actively touches only $\sim$0.5 MiB of cache lines. 
Most pages contain at least one hot object but remain mostly unused, creating vast regions of theoretically reclaimable memory that current systems cannot efficiently recover. 
This leads to our \textbf{observation \#1: improving Page Utilization increases reclaimable memory by reducing the number of pages needed to serve skewed workloads}.

Poor Page Utilization significantly impacts CPU performance by forcing hot objects to scatter across virtual address space.
Processors must perform frequent TLB lookups and page table walks even for cache-friendly workloads, with TLB misses requiring 150-600 cycles compared to 4-cycle hits. 
At Google, 11\% of fleet CPU cycles are consumed by dTLB load misses~\cite{tcmalloc}. 
While transparent huge pages reduce TLB pressure, applying them indiscriminately increases memory footprint by up to 69\%~\cite{tlb-improve-2}. 
Object grouping creates dense regions ideal for targeted huge page promotion, preserving TLB efficiency without memory bloat. 
This leads to our \textbf{observation \#2: improving Page Utilization enables targeted huge page promotion, preserving CPU cycles without sacrificing reclaimable memory}.

Poor address space layout forces datacenter infrastructure waste through memory overprovisioning and CPU underutilization.
Operators must spread jobs with small working sets but large allocation footprints across multiple machines to avoid CPU stranding~\cite{google-borg}, even when actual memory needs are much smaller than allocated. 
With DRAM accounting for 50\% of server costs~\cite{pond} and producing 12x more emissions per bit than SSDs~\cite{fairyWren}, reorganizing the address space by object access patterns creates a foundation for both memory efficiency and CPU optimization. This leads to our \textbf{observation \#3: better address space layout enables both efficient huge page usage and memory reclamation, allowing stable skewed workloads to run on fewer machines for more cost-effective and environmentally sustainable datacenters}.

\vspace{-2mm}
\section{Workload-Optimized Address Space}

The semantic gap between object-level allocation and page-level management demands a new approach to organizing the virtual address space. 
Instead of treating memory layout as a static side-effect of allocation, a system can dynamically engineer it to match workload behavior. This section outlines the core principles for designing such a system.

\subsection{Enabling Object Mobility}
\label{object-mobility}
The first challenge in reorganizing the address space is that applications written in unmanaged languages like C++ assume object addresses are fixed after allocation. 
To enable dynamic reorganization, a system must be able to move objects without violating this stability assumption from the application's perspective. 
The most practical way to achieve this is to \textbf{focus on pointer-based data structures,} where data is accessed through pointers rather than by arithmetic offsets. By intercepting pointer dereferences, a system can track and update pointers when an object moves, making the relocation transparent to the application. 
This approach requires no code modifications from the developer, as it works directly with objects allocated via \texttt{new} or \texttt{malloc}.
 This approach must be safe in a highly concurrent environment, requiring a lock-free mechanism that guarantees correctness without acquiring coarse-grained locks. 
 This ensures forward progress at all times and compatibility with the full spectrum of concurrency models, from simple locking to sophisticated lock-free algorithms.
 
 Applying this principle is fundamentally harder in C++ than in managed environments like Java or .NET. Managed runtimes have built-in moving garbage collectors that provide the machinery to safely relocate objects~\cite{oor-oopsla-2004, halo-pldi-2006}. 
 They can pause application threads at well-defined "safe-points" to scan the heap and atomically update all references to a moved object. C++ offers no such infrastructure. 
 Therefore, the challenge is not the idea of moving objects, but introducing a mechanism to do so safely and concurrently in a non-cooperative environment. 
 A system must be able to relocate an object while other threads may be actively trying to access it, without a "stop-the-world" pause and without breaking language semantics.

\subsection{Grouping Objects by Access Intensity}
\label{runtime-object}
Once objects can be moved safely, the system needs a policy to guide their placement. Relying on static, allocation-time placement \cite{atmem, xmem, tcmalloc-hotcold} is insufficient, as individual objects follow their own unique access trajectories---some remain hot throughout their lifetime, others cool down quickly after initialization, and many transition between hot and cold states as application phases change. 
An effective approach is to \textbf{dynamically group objects based on their observed access intensity,} a technique explored in managed runtimes for improving CPU cache locality~\cite{oor-oopsla-2004, halo-pldi-2006}.
The system monitors which objects are accessed over time, allowing it to differentiate between hot (active) and cold (inactive) data at runtime. 
This allows the system to adapt to shifting hot sets and application phase changes without prior knowledge of the workload.

This grouping creates an address space layout that directly supports memory efficiency. Hot objects are consolidated into dense regions, making them ideal candidates for huge pages to improve TLB performance. 
Cold objects are clustered together into separate regions, creating uniformly cold pages that can be easily identified and reclaimed.
Crucially, the mechanism for tracking this activity must have very low overhead. 
Unlike heavyweight profilers or dynamic instrumentation tools, a production-ready solution requires a lightweight mechanism to monitor object accesses without impacting application performance. 
By grouping objects based on their actual usage, the system organizes the address space to reflect the workload's true temporal access patterns.

\subsection{Decoupling Layout from Reclamation}
\label{frontend-backend}
The final principle is to \textbf{decouple object-level placement from page-level reclamation} to best leverage existing OS mechanisms. 
This separation establishes a frontend system that organizes the virtual address space and a back-end policy that acts upon it. 
The front-end's responsibility is to group objects by access intensity, creating regions of uniformly hot or cold pages. 
It provides an address space layout that is affable for any reclamation policy to act upon, whether that back-end is the kernel's kswapd or a user-space agent like TMO.

This separation makes existing back-ends more effective even with simple page placement policies.
When a back-end is presented with a page from a cold region, it no longer has to guess whether the page contains hidden hot objects; the front-end's organization provides a strong guarantee of its temperature. 
This allows reclamation policies to act more decisively and reclaim more memory without risking performance degradation from unexpected page faults. 
This decoupled design enables independent innovation: front-ends can improve their tracking and placement algorithms, while back-ends can evolve to support new hardware like CXL memory \cite{demystifying-cxl, wiscsort, systematic-cxl, mess-cxl}, all without requiring changes to the other layer.





\section{HADES}
\begin{figure}[!t]
  \centering
  \includegraphics[width=0.484\textwidth]{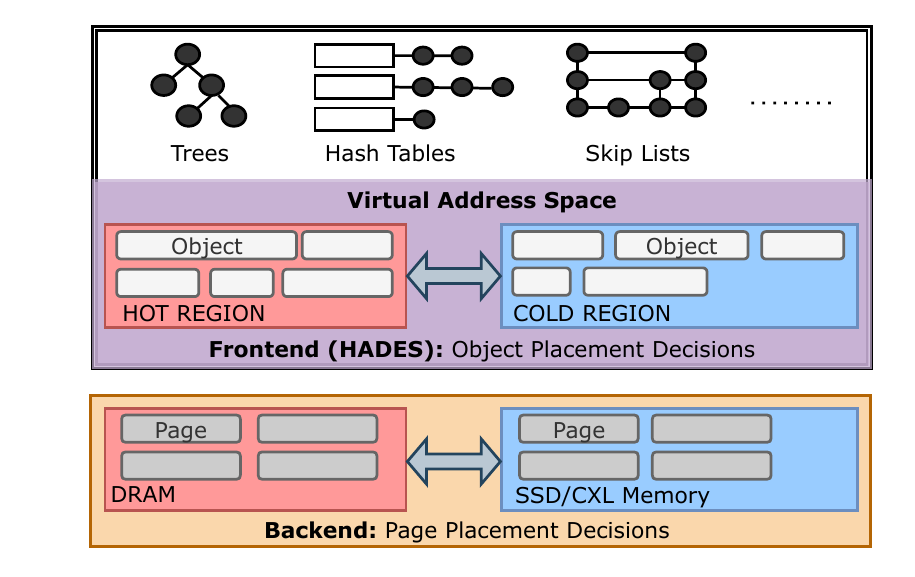}
  \caption{\textbf{HADES Overview.} 
  \small{As a front-end, HADES makes object placement decisions to organize the virtual address space into hot and cold regions. This enables any back-end to make more effective page placement decisions between DRAM and tiered storage.
  }}
  \label{fig:hades-overview}
\vspace{-4mm}
\end{figure}

HADES is a compiler-runtime co-design that implements the front-end role described in Section 3. As shown in Figure~\ref{fig:hades-overview}, it works seamlessly with existing page-level reclamation back-ends by providing them with an address space that is already organized by access activity. HADES monitors object access patterns and dynamically reorganizes the virtual address space, grouping recently accessed objects together while segregating rarely accessed ones. While HADES requires no changes to application logic, developers provide a one-time annotation to indicate which pointer-based data structures should be managed, enabling fine-grained optimization without altering program semantics.

\begin{figure}[!t]
  \centering
  \includegraphics[width=0.484\textwidth]{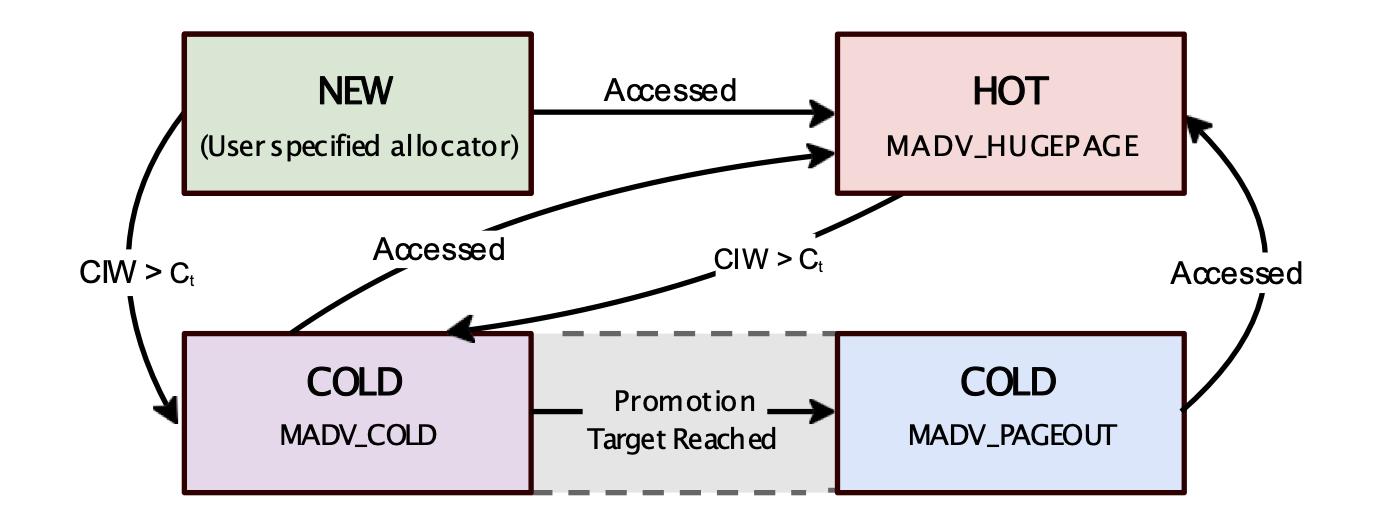}
  \caption{\textbf{Object Migration State Diagram.} 
  \small{An object starts in NEW heap and migrates to HOT or COLD based on access patterns and if its Consecutive Inactive Windows (CIW) is greater than a threshold ($C_t$), with heap properties adjusted through madvise flags.}}
  \label{fig:heap-state-diagram}
  \vspace{-4mm}
\end{figure}

\myparagraph{Tracking and Grouping Objects} To implement the principle of grouping objects by access intensity (\S\ref{runtime-object}), HADES first needs a low-overhead way to track access. Modern 64-bit processors leave high-order address bits unused, which HADES repurposes to store metadata directly within pointers~\cite{tagged-ptr}. It uses these "tagged pointers," or guides, to atomically set an access bit upon dereference, minimizing overhead by skipping the update to the access bit if it is already set. A runtime component, the Object Collector, periodically scans a sparse bitmap referencing all managed objects to read these access bits. Based on this information, it groups objects into three distinct heaps: a NEW heap for initial allocations, a HOT heap for frequently accessed objects (placed on huge pages), and a COLD heap for inactive objects targeted for reclamation (Figure~\ref{fig:heap-state-diagram}). A custom memory allocator ensures these heap regions are contiguous, enabling efficient madvise operations to form hugepages or page out the entire regions.

\myparagraph{Safe Concurrent Migration} To enable object mobility (\S\ref{object-mobility}), HADES must relocate objects without locks or application stalls. The key insight is to track an object's active use in real time, rather than its lifetime. HADES embeds a small \ARCfull{} (\ARC{}) in each guide's unused bits, which counts the number of threads currently executing within a function that accesses the object. To manage these counts efficiently, compiler-inserted scope guards automatically increment the \ARC{} at the start of a function and decrement it upon exit. This entire tracking mechanism is selectively activated using an epoch-based protocol only when migration is occurring, eliminating overhead during normal execution~\cite{faster,fraser}. During a migration window, the system can safely move any object with an \ARC{} of zero using an optimistic approach.

\myparagraph{Adaptive Workload Response} Fixed threshold migration policies are brittle and cannot adapt to changing workloads. HADES therefore employs a dynamic feedback loop to adjust its reclamation aggressiveness. It monitors the "promotion rate"~\cite{zswap} -— the rate at which applications access data from the COLD heap, which serves as a proxy for page-fault pressure. If this rate exceeds a configurable performance target (e.g., 1\%), it suggests the system is being too aggressive.
In response, we adapt concepts from TCP congestion control~\cite{tcp-aimd} to introduce a multiplicative increase/additive decrease (MIAD) strategy, which makes it harder for objects to be demoted. 
Initially, cold pages are marked with \texttt{MADV\_COLD} for reactive reclamation; the system only transitions to proactive \texttt{MADV\_PAGEOUT} once the promotion rate is safely below the target.

\myparagraph{Backend Integration} HADES validates the principle of decoupling layout from reclamation (\S\ref{frontend-backend}) by enhancing existing page-level mechanisms without modifying them. The Object Collector produces uniformly cold regions that kernel mechanisms like kswapd, TPP \cite{tpp}, or HeMem \cite{hemem}  can migrate with high confidence. 
As shown in Sec.~\ref{eval}, this architectural separation allows backends to operate more effectively, validating our proposed paradigm where a frontend provides object-level intelligence and the backend manages memory using its established policies.

\section{Evaluation}
\label{eval}

We evaluate HADES as a frontend that reorganizes address spaces to work with existing memory management backends. Our experiments demonstrate that object-level temporal tracking eliminates hotness fragmentation while enabling unmodified page-level reclamation systems to achieve both memory savings and performance preservation.


\myparagraph{Setup} We evaluate HADES' effectiveness using 
ten popular pointer-based data structures with diverse concurrency mechanisms (Table~\ref{tab:ds}), borrowed from ASCYLIB~\cite{ascylib}. For consistent testing, we develop CrestDB, a lightweight concurrent key-value store that can use any of these structures as its backend. Experiments were conducted on an Intel Xeon Gold 5218 CPU (16 cores), 32GB DRAM, and a 512GB Intel P4800x SSD for swap on Ubuntu 22.04. Each test ran six server threads and six client threads to stress concurrent access patterns.

\begin{table}
\centering
\scriptsize
\begin{tabular}{lll}
\hline
\textbf{Structure} & \textbf{Concurrency Control} & \textbf{Used In} \\
\hline
HashTable Harris \cite{harris} & Lock-free algorithm & NGINX \\
HashTable Pugh \cite{pugh} & Fine-grained r/w lock & Redis, Memcached \\
HashTable Java CHM \cite{chm} & Segmented bucket locks & Linux kernel, HAProxy \\
\hline
SkipList Coarse & Global lock & LevelDB/RocksDB \\
SkipList Fraser \cite{fraser} & Lock-free algorithm & Redis Sorted Sets \\
SkipList Herlihy \cite{herlihy} & Optimistic fine-grained & Cassandra, CockroachDB \\
\hline
B+Tree Coarse & Global lock & SAP HANA \\
B+Tree OCC & OCC w/ epoch reclaim & VoltDB index \\
MassTree \cite{masstree} & OCC + RCU & LMDB \\
\hline
Adaptive Radix Tree \cite{art} & Fine-grained r/w lock & DuckDB, PostgreSQL \\
\hline
\end{tabular}
\caption{Concurrent data structures evaluated with HADES}
\label{tab:ds}
\vspace{-10mm}
\end{table}

\begin{figure*}[t]
    \centering
    \begin{minipage}[t]{0.34\textwidth}
        \vspace{0pt}
        \centering
        \includegraphics[width=\textwidth,height=4cm,keepaspectratio]{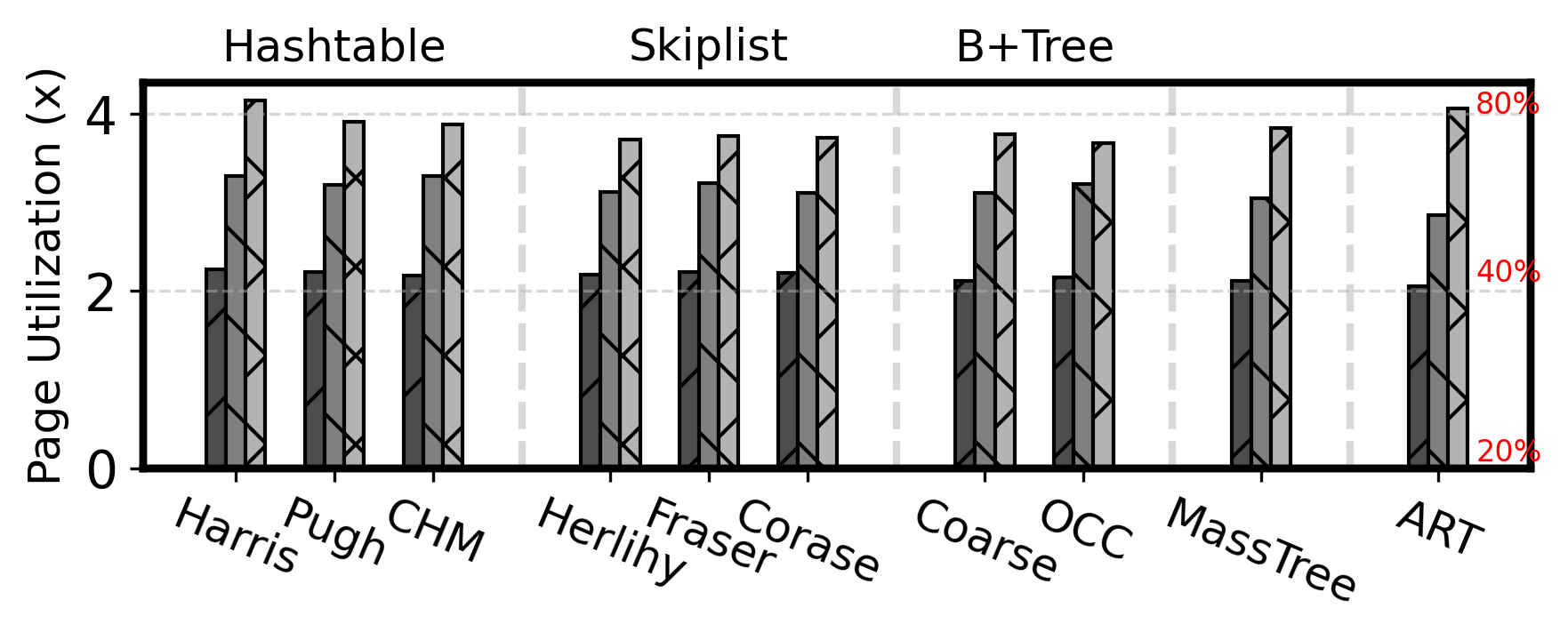}
        \caption*{\centering \textbf{(a) Page Utilization Improvement.} \textmd{\small{After HADES migrates objects, utilization increases for read-heavy workloads; by up to 84\% and 40-60\% for workloads A and B.}}}
        \vspace{-3mm}
        \label{fig:pu-figa}
    \end{minipage}%
    \hfill
    \begin{minipage}[t]{0.34\textwidth}
        \vspace{0pt}
        \centering
        \includegraphics[width=\textwidth,height=4cm,keepaspectratio]{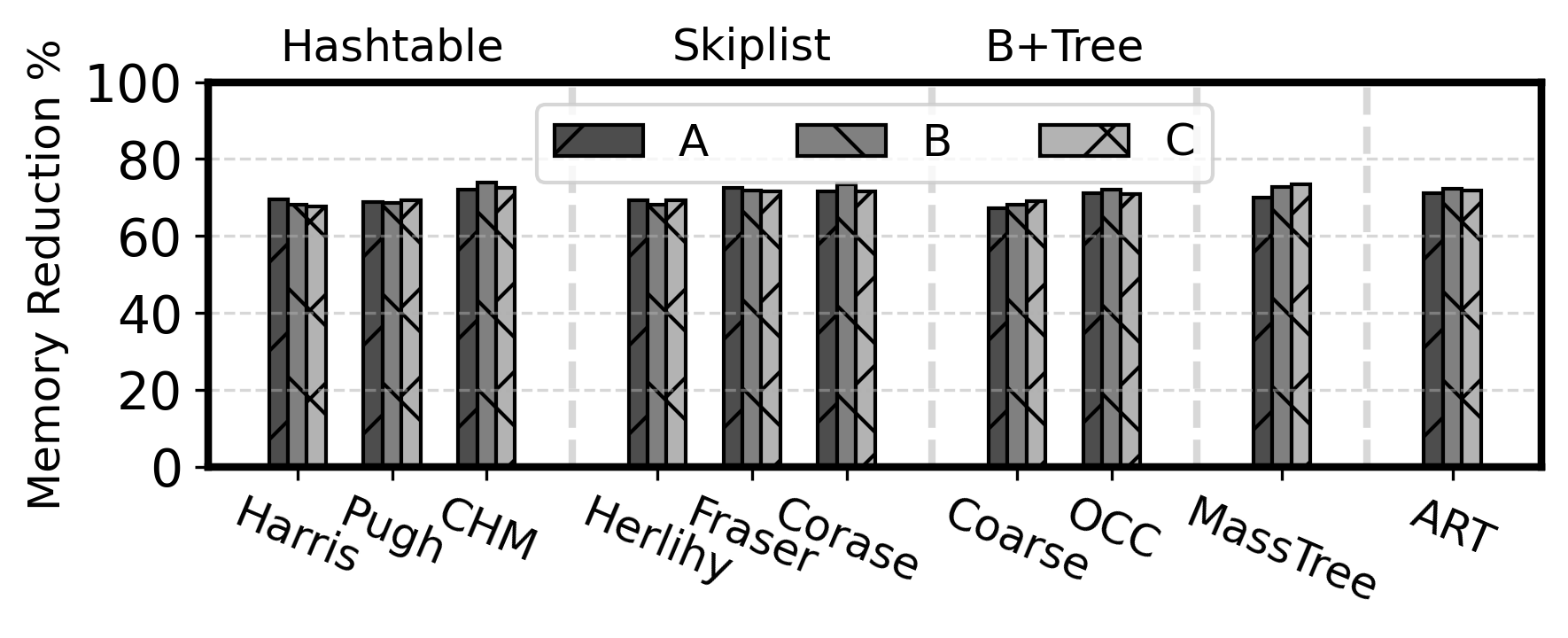}
        \caption*{\centering \textbf{(b) Memory Usage.} \textmd{\small{HADES identifies cold objects and reclaims their pages, reducing memory usage by up to 70\% for all three workloads.}}}
        \label{fig:mem-figb}
    \end{minipage}%
    \hfill
    \begin{minipage}[t]{0.31\textwidth}
        \vspace{0pt}
        \centering
        \includegraphics[width=\textwidth,height=4cm,keepaspectratio]{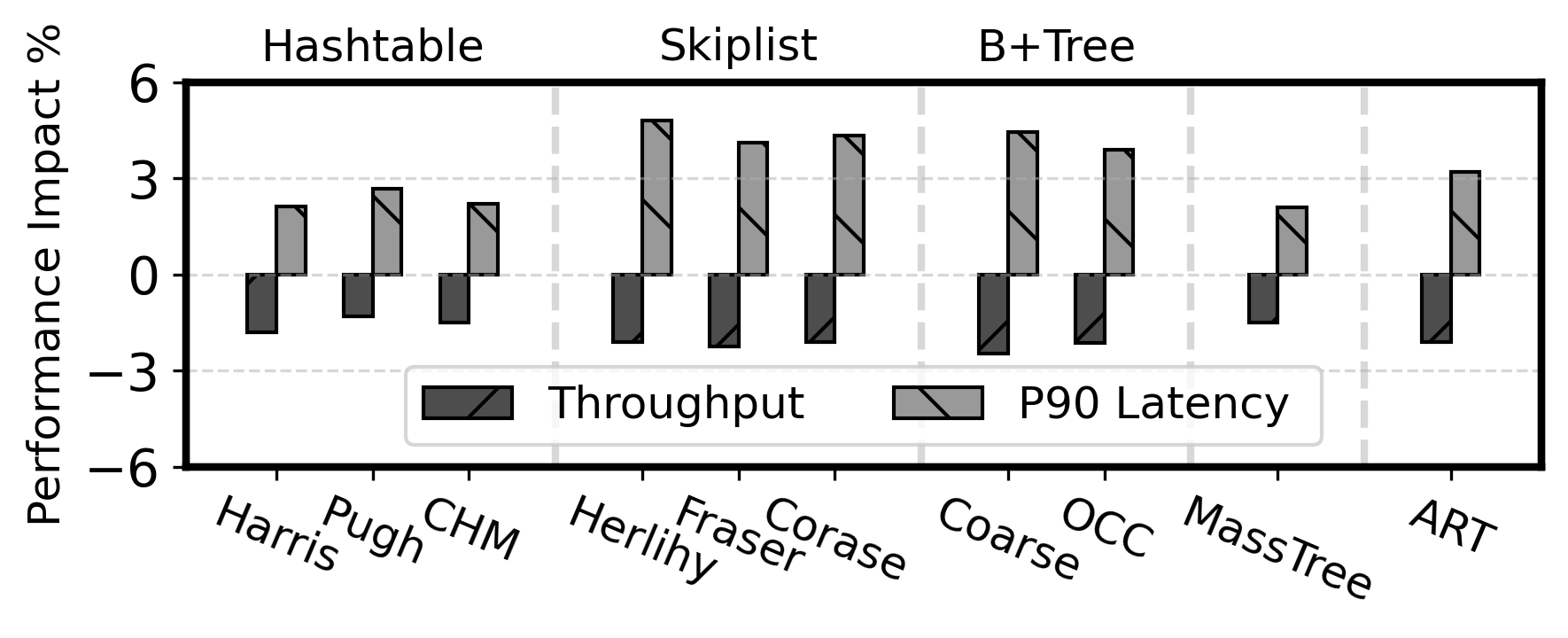}
        \caption*{\centering \textbf{(c) Performance Impact Run-A.} \textmd{\small{HADES incurs throughput and latency degradation due to \ARC{} instrumentation, varying by data structure complexity.}}}
        \label{fig:perf-figc}
    \end{minipage}
    \caption{\textbf{CrestDB with YCSB.} \textmd{\small{HADES effects on page utilization, memory usage, and performance across YCSB workloads A (50\% writes), B (5\% writes), and C (read-only) under a zipfian distribution. }}}
    \vspace{-4mm}
    \label{fig:ycsb-overview}
\end{figure*}

\subsection{Frontend Effectiveness on YCSB Workloads}

We tested our key-value interface against three YCSB workloads using zipfian distributions that scatter hot keys throughout the entire key space, creating realistic access skew unlike YCSB's default hot key concentration patterns~\cite{hotring, benchmark-rocksdb-fb, twitter-traces}. Each experiment loads 10M KV pairs with 30B keys and 1024B values to demonstrate address space optimization effectiveness across diverse data structure implementations.

HADES increases page utilization from 18-20\% to substantially higher levels through object grouping across all data structure types. Once the system completes its initial object classification phase, it achieves 2x improvement for workload A, 3x for workload B, and 4x for workload C (Figure~\ref{fig:ycsb-overview}(a)). Read-only workload C reaches 80\% page utilization as HADES migrates hot objects to dedicated regions without interference from new allocations, while update-heavy workloads show lower improvements because new value allocations initially appear in the NEW heap. The consistent improvement across all ten data structures—from simple hash tables to complex B+Trees with different concurrency mechanisms—confirms that object-level tracking eliminates hotness fragmentation regardless of the underlying index implementation or synchronization approach.

The frontend identifies and reclaims cold memory effectively across all workloads and data structure types. Figure~\ref{fig:ycsb-overview}(b) shows HADES reduces memory usage by up to 70\% through object-level cold identification and heap organization. When promotion rates reach target levels indicating accurate cold classification, HADES transitions from marking pages with \texttt{MADV\_COLD} to issuing \texttt{MADV\_PAGEOUT} for proactive reclamation. This object-level reorganization enables precise working set identification that creates uniformly cold pages suitable for safe reclamation without performance risks.

HADES introduces 2.5\% average throughput reduction and 5\% latency increase from tracking instrumentation overhead. Performance impact varies by data structure complexity, with hash tables showing lower overhead than skiplists and B+Trees due to differences in traversal patterns and key comparison requirements (Figure~\ref{fig:ycsb-overview}(c)). Access bit operations consume 4-5 ns (comparable to L1 cache hits), while the primary overhead occurs during scope guard operations requiring O(logN) complexity for tracking first-time object observations. The modest overhead across diverse concurrency mechanisms demonstrates that object-level tracking remains practical regardless of the underlying synchronization approach.

\begin{figure}[t]
    \centering
    \includegraphics[width=0.49\textwidth]{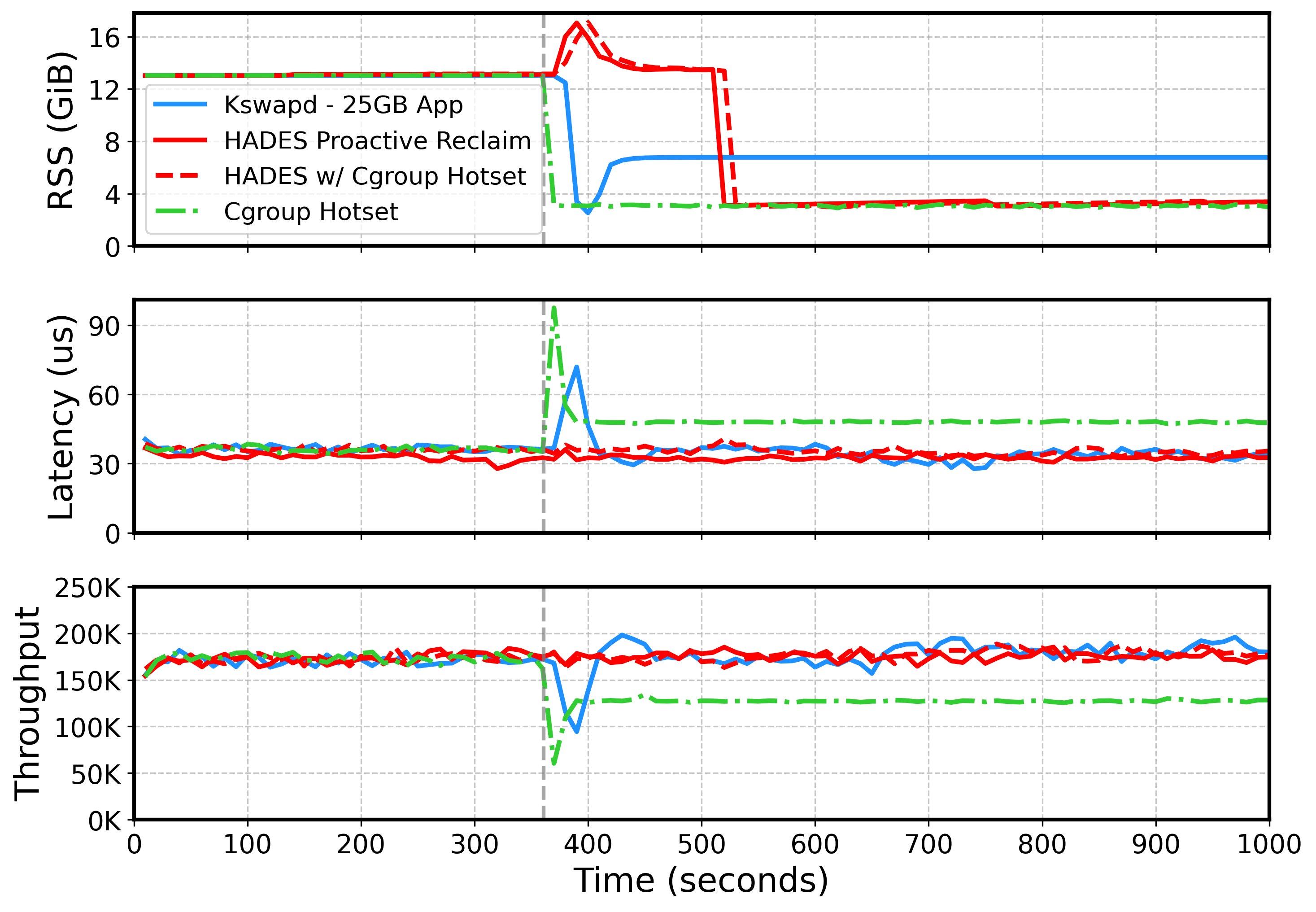}
    \vspace*{-4mm}
    \caption{
        \textbf{CrestDB vs. Baselines on YCSB-C.}
        \textmd{
        \small{HADES resolves the trade-off between performance and memory savings. Standard backends must either sacrifice performance to save memory (Cgroup hotset) or sacrifice memory savings to preserve performance (kswapd). By providing an organized address space, HADES enables both reactive (HADES w/ Cgroup hotspot) and proactive (HADES Proactive) reclamation to achieve maximum memory savings with no performance degradation.}
        }
    }
    \label{fig:baseline}
    \vspace{-6mm}
\end{figure}

\subsection{Backend Integration Validation}

We demonstrate that HADES enables existing backends to achieve both memory savings and performance preservation by running YCSB-C with a 12GB footprint while actively accessing only $\sim$4GB. Traditional systems force operators to choose between aggressive reclamation that degrades performance and conservative approaches that waste memory. HADES eliminates this trade-off by providing backends with uniformly cold pages that reclaim without affecting hot data.

Figure~\ref{fig:baseline} reveals the fundamental limitation of page-level backends operating on poorly organized address spaces. The memory-saving-first approach (green line) sets cgroup limits to 4GB, achieving target memory usage but causing 50\% latency increase and 30\% throughput degradation from aggressive reclamation of pages containing hot objects. The performance-first approach (blue line) creates pressure through background applications (allocates 25 GB), preserving performance but achieving poor memory savings with usage remaining around 6GB rather than the 4GB target. Both backends face the same constraint: page-level granularity cannot distinguish truly hot pages from those containing scattered hot objects.

HADES transforms the same backends from facing painful trade-offs to achieving both objectives simultaneously. The reorganized address space enables \texttt{kswapd} to reduce memory usage to 4GB without performance degradation (dashed red line) because the COLD heap contains uniformly cold objects safe for reclamation. Proactive reclamation through \texttt{madvise} achieves identical results (solid red line), demonstrating that multiple backend approaches benefit from frontend organization. The backend integration validates that object-level intelligence creates conditions where page-level mechanisms operate effectively without modification.
\vspace{-3mm}
\section{Limitations}

HADES optimizes the placement of managed objects in the virtual address space to achieve better page utilization, which enables proactive memory reclamation and improved TLB efficiency. Despite these advantages, HADES faces several limitations that affect its applicability across different scenarios.

\begin{itemize}
    \item \textbf{Lack of pointer stability:} HADES invalidates cached pointers by moving objects across memory, requiring users to query keys and values each time rather than caching pointers. This mirrors constraints in Abseil containers \cite{abseil} and STL structures like \texttt{std::vector} and \texttt{std::deque}, which similarly don't guarantee pointer stability.
    
    \item \textbf{Unique Object Ownership:} When a pointer is annotated for HADES management, users implicitly assert that it has exclusive access control over the object. This unique ownership model, akin to \texttt{std::unique\_ptr} semantics, is essential for safe object migration, as HADES only updates the address via the annotated pointer. This model aligns with the internal object management within the kinds of concurrent data structures evaluated (see Table \ref{tab:ds}), which are fundamental to various systems like databases, in-memory caches, and web servers. These structures naturally create unique ownership paths for their elements or nodes. Objects shared and accessed through multiple aliases cannot be safely managed by HADES.
    
    \item \textbf{Incompatibility with pointer arithmetic:} HADES places objects across multiple heaps without maintaining contiguous placement, eliminating implicit ordering guarantees and making it unsuitable for data structures like arrays and matrices that require contiguous memory.
    
    \item \textbf{Language support restrictions:} HADES only works with languages supporting operator overloading (C++, Rust) to intercept access during pointer dereferencing. Languages without this capability (Go, Java) cannot implement this approach, though they might enable direct object management through garbage collection.
    
    \item \textbf{Requirement for explicit annotations:}  Users must designate which objects HADES manages, which is straightforward in key-value stores but perhaps challenging in more complex scenarios. Determining automatic management candidacy falls outside this work's scope.
\end{itemize}

\vspace{-2mm}
\section{Related Work}
\textbf{Object-Level Management.} Recent works like AIFM \cite{aifm} and MIRA \cite{mira} operate at object granularity but focus exclusively on far-memory over RDMA, requiring direct hardware access that limits production adoption. Alaska \cite{handle} uses handle-based indirection to reduce RSS through heap compaction, addressing fragmentation reactively without object hotness classification. HADES takes a fundamentally different approach by proactively reshaping the virtual address space through temporal access tracking, creating tiering-friendly object clusters that bridge the gap between application-level object access and page-level OS memory management. This organization of objects across specialized heaps based on access frequency enables HADES to work effectively with existing OS reclamation mechanisms while adapting to changing workload characteristics.

\textbf{Runtime vs. Allocation-Time Placement.} Allocation-time hinting approaches \cite{xmem, atmem, data-spatial-locality, tcmalloc, llvm-pgho} fail to capture objects transitioning between hot and cold states or distinguish between objects from the same allocation site with different access patterns. Systems like PGHO~\cite{llvm-pgho} can apply hints automatically based on profile data for allocators like TCMalloc \cite{tcmalloc}, but still make placement decisions only at allocation time. These instrumentation-based profile collection solutions are too slow for production workloads and rely on representative workloads that are often unavailable. HADES instead tracks access patterns at runtime, enabling migration based on actual usage rather than static predictions.

\textbf{Page-Level Optimizations.} In contrast to HADES's object-level address space reorganization, several systems optimize memory tiering and efficiency at the page level. 
For instance, Johnny Cache \cite{jc} manipulates physical page allocation to minimize address conflicts in the hardware sets of direct-mapped DRAM caches. Similarly, Memstrata \cite{memstrata} manages host physical page mappings in virtualized environments to isolate tenants and optimize performance within Intel Flat Memory Mode, where hardware tiers cachelines between local DRAM and CXL memory. These systems improve how pages interact with the underlying cacheline-granular hardware, but they treat the page contents as opaque. HADES's approach is orthogonal, as it ensures the cache lines within each virtual page are more uniformly hot or cold, making the hardware's tiering decisions more effective.

Approaches also exist for page-level decisions about tiering and page size. HawkEye~\cite{hawkeye} improves huge page management through fine-grained page-level access tracking for better TLB utilization. Memtis~\cite{memtis} dynamically determines page tier placement and page size in heterogeneous memory systems. Based on the access skew within huge pages detected via hardware sampling, Memtis decides whether to split a huge page into smaller base pages, migrating only the hot subpages to the fast tier, thus balancing tiering benefits against address translation costs. While Memtis optimizes page sizes and placement, it still operates at the page level. If a base page still contains a mix of hot and cold objects, Memtis cannot separate them. HADES, by organizing objects, ensures that pages are more likely to be homogeneously hot or cold, potentially improving the effectiveness of systems like Memtis, Johnny Cache, and Memstrata.
\vspace{-2mm}
\section{Conclusion} 
We demonstrate that the key to efficient memory tiering in non-managed languages lies not in altering the OS, but in engineering the application's virtual address space to be OS-friendly.
We introduced a decoupled frontend/backend model where an object-level frontend dynamically reorganizes the address space to create uniformly hot and cold regions, enabling any standard page-level backend to reclaim memory far more effectively. 
HADES realizes this vision by introducing novel, lock-free techniques to safely migrate objects in a concurrent environment. By resolving hotness fragmentation at its source, HADES allows standard backends to reduce memory usage by up to 70\% without the performance trade-offs that have made aggressive memory tiering impractical.

\section*{Acknowledgments}
We thank David Culler, Lilian Tsai, Qian Ge, Teresa Johnson, colleagues in SystemsResearch@Google, the anonymous reviewers, and our shepherd, Antonio Barbalace, for valuable feedback on earlier drafts of this manuscript.


\bibliographystyle{ACM-Reference-Format}
\bibliography{ref}
\end{document}